\documentclass[sigconf]{acmart}
\usepackage{tabularray}
\usepackage{soul}

\AtBeginDocument{%
  }

\copyrightyear{2026}
\acmYear{2026}
\setcopyright{cc}
\setcctype{by}
\acmConference[CHIIR '26]{2026 ACM SIGIR Conference on Human Information Interaction and Retrieval}{March 22--26, 2026}{Seattle, WA, USA}
\acmBooktitle{2026 ACM SIGIR Conference on Human Information Interaction and Retrieval (CHIIR '26), March 22--26, 2026, Seattle, WA, USA}
\acmDOI{10.1145/3786304.3787863}
\acmISBN{979-8-4007-2414-5/2026/03}

\begin{document}

\title{Rules, Resources, and Restrictions: A Taxonomy of Task-Based Information Request Intents}

\author{Melanie A. Kilian}
\orcid{https://orcid.org/0000-0003-0182-926X}
\affiliation{%
  \institution{University of Regensburg}
  \city{Regensburg}
  \country{Germany}}
\email{melanie.kilian@ur.de}

\author{David Elsweiler}
\orcid{https://orcid.org/0000-0002-5791-0641}
\affiliation{%
  \institution{University of Regensburg}
  \city{Regensburg}
  \country{Germany}}
\email{david.elsweiler@ur.de}

\renewcommand{\shortauthors}{Melanie A. Kilian and David Elsweiler}

\begin{abstract}
Understanding and classifying query intents can improve retrieval effectiveness by helping align search results with the motivations behind user queries. However, existing intent taxonomies are typically derived from system log data and capture mostly isolated information needs, while the broader task context often remains unaddressed.
This limitation becomes increasingly relevant as interactions with Large Language Models (LLMs) expand user expectations from simple query answering toward comprehensive task support, for example, with purchasing decisions or in travel planning. At the same time, current LLMs still struggle to fully interpret complex and multifaceted tasks. To address this gap, we argue for a stronger task-based perspective on query intent.
Drawing on a grounded-theory-based interview study with airport information clerks, we present a taxonomy of task-based information request intents that bridges the gap between traditional query-focused approaches and the emerging demands of AI-driven task-oriented search.
\end{abstract}



\maketitle

\section{Introduction}
\label{sec:intro}

Understanding why people seek information is essential for designing information search systems that support their users effectively \citep[p. 2436]{shah2020tutorial}. Research on \textit{query intent understanding and classification} (QIU\&C) has therefore aimed to derive the purpose behind search queries and assign them to taxonomies of user goals \cite[p. 3]{deng2020intro}. Such taxonomies can improve retrieval effectiveness \cite[p. 15]{guo2020query} and user satisfaction \cite{su2018user,bodonhelyi2024user}.
To date, numerous classifications have been proposed. These range from Broder’s influential taxonomy of general-purpose Web search \cite{broder2002taxonomy} to more domain-specific frameworks in areas such as e-commerce, (multi-)media search, and legal case retrieval \cite{yang2024bespoke,xie2018why,shao2024intent}. Related work has also extended to various request types, from factoid questions 
\cite{yin2010building,meier2021towards} to opinion-seeking inquiries \cite{bolotova2022nonfactoid,verberne2006data}, and diverse search modalities, such as question answering \cite{chen2012understanding,bolotova2022nonfactoid} and (AI-driven) conversational search \cite{qu2018analyzing,trippas2024what}. 
    
While this research has provided valuable insights, most existing taxonomies remain tied to isolated information needs and are typically built from system log data, such as query and click logs \cite[p. 36]{guo2020query}. These resources are limited: Queries are often ambiguous, context-poor, and ill-suited to capture the user’s higher-level task\footnote{We use "tasks" in a broad sense to include all the motivating reasons behind why people seek information, such as duties, interests, passions, or goals related to work, daily life, hobbies, or leisure activities. This reflects the view of tasks that prevails in task-based IR \cite[][p. 117]{belkin2013task}.} \cite{baeza2006intention}.
These problems also hold for AI-driven chat interfaces, where requests can still suffer from ambiguity and lack the broader task context \cite[p. 2704]{keluskar2024do,trippas2024what}.

At the same time, the rise of Large Language Models (LLMs) has shifted user expectations and system capabilities. Interactions increasingly extend beyond fact retrieval toward more complex use cases, such as decision making (e.g., choosing a product to buy) and task execution (e.g., planning a trip). To meet these needs, research must adopt a stronger task-based lens \cite{shah2025LLMs,shah2025from,he2025plan},
prioritizing high-level work tasks over scattered information requests. Yet current LLMs continue to struggle with fully understanding such complex tasks, for example when temporal dependencies between sub-tasks must be specified \cite{yuan2024tasklama}.

We propose that additional qualitative resources are needed to complement log-based approaches and to capture task-based request intents more accurately. In this paper, we draw on qualitative expert interviews \cite{bogner2009theory} with information clerks who, through their daily practice, develop long-term insights into both users’ information needs (\textit{the user lens}) and the underlying tasks, procedures, and regulations shaping those needs (\textit{the task lens}). By analyzing these insights, \textbf{we contribute a taxonomy of task-based information request intents that bridges the gap between established query-focused classifications and the emerging demands of (AI-driven) task-oriented search}.

Our grounded theory (GT) based interview study gives insights into
the experiences of eight airport information desk workers, spanning 120
years of professional engagement with task support. 
The airport context has proven to be a place where people perform a diverse range of tasks that motivate information seeking. These include complex "work" tasks around air travel, purely hedonic activities, and combinations of these.

Additionally, airport visitors await fixed workflows, tight schedules, and diverse distances to navigate through. This makes the setting also rich of context conditions and lets visitors often turn to on-site information desks for help.
All this establishes the airport as a particularly suitable context for building a multifaceted taxonomy of task-based request intents. This diversity of tasks and context conditions found at airports makes our taxonomy moreover likely to be transferable to other task-based contexts.
The GT methods used are intended to contribute to this by raising the level of abstraction during taxonomy building (e.g., constant comparisons \cite[][p. 187]{charmaz2006constructing} and theoretical comparisons \cite[][pp. 94–95]{corbin2015basics}).

\section{Related Work}

Aiming for a taxonomy of task-based request intents touches two important research areas in contemporary (interactive) information retrieval, (I)IR for short: (1) \textit{query intent classification} and (2) \textit{task-based information retrieval (IR)}. 
In the following, we summarize work in these two research areas to demonstrate the individual contributions made by these research branches, show where they currently come to their limits, and how combining both branches, as pursued in this paper, may exceed those limits and promote the development of more effective task support systems. 

\subsection{Research on Query Intent Classification}

Query intent classification is known under varying terms depending on the specific research sub-field it is practiced. Besides "query intent", terms, such as "search intent", "question intent", and "user goal", are also common. This already gives an indication of how wide this research field is. It is therefore not surprising that numerous taxonomies have been proposed to date.

Broder’s early taxonomy of Web search may be the one most known in each sub-field \cite{broder2002taxonomy}. It differentiates between informational, navigational, and transactional intents behind searches. Broder's taxonomy has been widely expanded upon, resulting in more fine-grained classifications for general-purpose Web search \cite{rose2004understanding,baeza2006intention,jansen2007determining,cambazoglu2021intent}.
Beyond general taxonomies, there are also numerous domain-specific classifications, spanning areas such as e-commerce \cite{dai2006detecting,yang2024bespoke,ahmadvand2020jointmap}, (multi-)media search \cite{xie2018why,kofler2016user}, and legal case retrieval \cite{shao2024intent}. Further specifications in taxonomies concern types of requests, which resulted in classifications solely focusing on factoid questions (e.g., named-entity queries \cite{yin2010building} and known-item requests \cite{meier2021towards}) or non-factoid questions asking for opinions and experiences \cite{bolotova2022nonfactoid,verberne2006data}. Intent taxonomies have also been built for various search modalities, such as question answering \cite{chen2012understanding,bolotova2022nonfactoid} and (AI-driven) conversational search \cite{qu2018analyzing,trippas2024what}.
Taken together, these classification schemes have created a rich and multi-perspective pool of possible question intents behind searches, promoting a user-centered lens in information retrieval research. Additionally, intent taxonomies show practical relevance by improving retrieval effectiveness \cite[p. 15]{guo2020query} and user satisfaction \cite{su2018user,bodonhelyi2024user}.  
When it comes to the analytical perspective taken, a substantial portion of existing intent taxonomies are content-based, focusing on the expected type of answer content, such as "disease", "musical instrument", and "measurement". For broader applicability in general-purpose search, function-based taxonomies have been developed. These aim to classify questions based on the role or function of the expected answer. For example, Bu and colleagues' (2010) \cite{bu2010function} taxonomy for general question answering contains functional answer categories, such as "reason", "solution", and "definition".

Regradless of their specification focus or analytical perspective, most intent taxonomies are built upon system log data, such as query logs, click logs, and retrieved documents \cite[p. 36]{guo2020query}. This type of data makes it possible to collect a vast number of questions. However, these resources often lack context, as they are tied to isolated information needs \cite{belkin2013task}. This makes it difficult to capture the higher-level task motivating the user to engage in information search. This also holds for AI-driven chat interfaces, where requests can still be context-poor \cite[p. 2704]{keluskar2024do,trippas2024what}. Thus, intent taxonomies with general task-oriented answer categories, such as "required actions" and "available resources", have not been proposed thus far. This is particularly unfavorable in the case of emerging search modalities such as AI-driven chat, where users increasingly expect overall task support. The need for task-focused taxonomies has, for example, already been recognized by Shah et al. \cite[][p. 34:2]{shah2025LLMs}.

\vspace{-1mm}
\subsection{Research on Task-Based IR}

For some time now, Shah and various colleagues \cite{shah2023tak,shah2025from,shah2021bridging} have been calling for a renewed focus on tasks in (I)IR. They make the importance of the task lens for a more effective user-centered search clear, particularly pointing to the rise of (AI) driven search modalities and proactive information systems, their game-changing capabilities, and the associated expectations of users for these systems.
For some decades, the task lens has been widely recognized and promoted for its benefits: 
In the 1990s, there was still a strict differentiation between studies on information needs, seeking and use (INSU) and research in the field of IR \cite[][p. 577]{bystrom2004conceptions}, with INSU focusing more upon the broader context of human information activities and IR specializing in highly specific system-centric contexts of information searching and finding. Byström and colleagues \cite{bystrom2004conceptions} pointed out that task-based investigations of both INSU and IR represent an opportunity to merge the two fields. Tasks narrow the research object of INSU down to the more specific context of task performance while, at the same time, expanding the research focus of IR with the more general level of a work task prompting the search of a information system \cite[][p. 578]{bystrom2004conceptions}. Thus, the task concept establishes a conceptual bridge between the two research areas \cite[][p. 1059]{bystrom2005conceptual} and helps bring them together to advance more effective and user-friendly search experiences.
Tasks are currently the predominant way to conceptualize the motivations for information seeking \cite{bystrom2013modeling,li2008faceted,li2010exploration} and have proven to be valuable in many areas of information-centered research \cite{meyers2009diverse}. Tasks represent an observable entity as opposed to the vague notion of information needs \cite[][pp. 82--91]{case2016looking}. As such, they serve as fruitful units for analysis
in empirical studies \cite{bystrom2002work} and can produce contextualized and thus more realistic findings of practical relevance \cite{kumpulainen2017task}.
Several task frameworks \cite[e.g.,][]{bystrom2005conceptual,toms2019information,soufan2021untangling,shah2023tak} describe how closely tasks relate to information seeking (e.g., from activities to (work) tasks to search tasks).
However, task-based IR in the sense of systems that support task completion by guiding users through different task stages remains still a future prospect with only small steps that have already been taken (e.g., context-sensitive search features \cite{shah2023tak}).
One further move toward task assistance may be to integrate the users' search intents during complex real-world task performances (the user lens) with the underlying tasks, procedures, and regulations shaping those user intents (the task lens). This requires a task-based analysis of information request intents and therefore context-rich data on user requests. 
Drawing on qualitative expert interviews with airport information clerks, we have taken on this challenge and propose a taxonomy of task-based information request intents.

\section{Methodology}

\subsection{General Methodological Approach}

We conducted a qualitative expert interview study \cite{bogner2009theory,charmaz2012qualitative} as part of a larger research project that explores information-related activities in airports through grounded theory methodology (GTM) \cite{corbin2015basics,charmaz2006constructing}. This paper addresses one of the guiding research questions of this project, viz., \textit{What various information needs do airport visitors have? And what associated tasks do they pursue?}, by developing a taxonomy of task-based information request intents.

In addressing these research questions, expert interviews \cite{bogner2009theory} with airport information desk workers have proven to be an especially suitable method. The information clerks' perspective covers both intensive knowledge on people's information needs and action goals (\textit{the user lens}) and an expert understanding of the tasks, procedures, and regulations, as well as the conditions and constraints existing in the specific information-seeking context (\textit{the task lens}). This provides a realistic portrayal of inquirers' request intents and makes it also possible to match these with task-oriented constructs that may only be implicit to lay users or interviewees (if they are not going through this situation right now).

\subsection{Participants}

Interview participants were selected through purposive sampling \cite{palys2008purposive} to gain insights into information needs and tasks of diverse types of inquirers. We regard information desk workers in international airports with different terminal layouts as suitable candidates. Such airports feature diverse visitors and contextual conditions which contribute to making our findings more applicable to other task contexts. Munich Airport in Germany falls under this category. Moreover, Munich Airport information desk workers support visitors across diverse communication channels (e.g., face to face, telephone, and social media) and in more than 15 languages (e.g., English, Hindi, and Spanish).

Access to this group was obtained in cooperation with Munich Airport. We first recruited interviewees via a gatekeeper \cite[p. 2]{jensen2008access}, their immediate supervisor. Later, to avoid gatekeeper bias \cite[p. 180]{temple2008cross}, we also directly approached interested workers on site \cite{thomas2007process}. This was done with their supervisor's approval. Only one of the workers declined to participate. Participation was uncompensated, but interviews were conducted during paid work hours. While recruiting, we aimed to diversify the sample based on different properties, such as age, professional experience, and the communication channels used by the expert to support visitors. 

Our sampling approach has resulted in a heterogeneous sample of information desk workers. It reflects the diversity of the experts in their professional experience, age, cultural background, and in the languages and communication channels they use to support visitors (see \href{https://asistdl.onlinelibrary.wiley.com/cms/asset/8a709386-afab-4798-b2c2-e7ffb22a6cd7/asi70010-fig-0001-m.jpg}{Figure 1} in \cite{kilian2025right} for more details). The sample includes eight women who together have more than 120 years of experience in information desk work.
In Section \ref{sec:findings}, we use pseudonyms to refer to individual participants.

\subsection{Methods for Data Collection}

The interviews were held at the participants’ workplace, in a private room with only the researcher and the interviewee present or at the participant’s desk in an open-plan office, where the interviewee simultaneously answered incoming information requests.

We combined the theory-generating expert interview method \cite{bogner2009theory} with the in-depth (grounded theory) interviewing approach \cite{charmaz2012qualitative} to question the information desk workers. Only open-ended questions, such as “Please share your experiences during today's shift” and “What requests do airport visitors make?,” guided the interviews.
To probe interviewees' reports without producing bias, specific questions, for example, “Do inexperienced travelers' requests differ from those of experienced travelers?,” were asked based on statements the interviewee had said before, for example,  “You just told me about requests from experienced travelers. What can you tell me about inexperienced travelers?” Many interview questions ultimately revolved around airport visitors' information requests and their associated tasks and information seeking behaviors.
In this non-directive way, we also challenged interviewees to produce diverse 
accounts of information requests. We encouraged them to differentiate information requests according to various factors, such as the communication channel used, the inquirers' age, airport experience, and cultural background, and the task motivating the requests (e.g., departing, picking arriving passengers up, or doing airport sightseeing). In this context, the interviewees also brought up other factors that we had not considered, such as differences in requests depending on the season.

Eight interviews were conducted over a 2-month period, between December 2016 and January 2017.
We audio recorded all interviews and made sure that ethical standards regarding informed consent and data collection were met.
The number of interviews and the amount of interview data (overall approx. 621 min) are in line with other studies using expert interviews \cite[e.g.,][]{duff2002accidentally, mccay2015investigating,siren2023crossroads}.\footnote{Experts are not individuals reporting on a single case. Our participants have handled many thousand cases over decades and can give accounts of diverse cases as the sample includes experts with varying intercultural experiences and experiences with varying communication contexts (e.g., interactions on site, via telephone, email, or social networks). The long and detailed nature of the interviews (overall approx. 621 min.) provides moreover deep insights into these multiple, diverse situations.}
Data collection ended when “theoretical sufficiency” \cite[][pp. 213 and 215]{charmaz2025constructing} was considered reached.

\subsection{Taxonomy Building: Methods for Data Analysis}

We analyzed the interview data by means of a grounded theory based, inductive approach. The data analysis followed the general course of the three GTM coding strategies: open, axial, and selective coding \cite[][p. 344]{corbin2015basics}. Memoing \cite[][p. 72]{charmaz2006constructing} and constant comparisons \cite[][p. 85]{corbin2015basics}  were used throughout the entire data analysis.

After transcribing the interviews verbatim, we used inductive open coding to identify every concrete mention of an information request and annotate these with labels describing the question intent.
Early on, it became clear that a task-context-based coding was essential to understand the intents behind the inquirers' information requests, for example, depending on the specific task context, not every \textit{"Where is ...?"} question calls for a statement of a location; sometimes inquirers expect way-finding directions instead. Thus, we coded all task descriptions in a further round of data analysis and used these task-focused codes to categorize requests more meaningfully.
In identifying mentions of information requests, we only coded statements that clearly indicate the question being asked by the information seeker (e.g., “I once had a Russian woman [ask me], ‘Is it safe [...] at the main station? [...] Can I go there alone?’”) 
In most cases, the experts phrased information requests in direct or indirect speech, often additionally adopting the inquirers' perspective by using first-person pronouns, such as "I" and "my" (see the question about the main station above).
We did not include vague or abstract descriptions of information requests (e.g., “We are currently receiving a lot of calls about lost luggage.”).
We also excluded help-seeking requests in which inquirers were interested in services rather than information (e.g., “A passenger came to me and said, ‘I would like you to cancel my hotel reservation at this or that hotel.’”).
This way, we identified \textbf{720 statements of information requests}.

As the analysis progressed, we used axial coding to identify connections within and between the labeled intent categories. This way, categories and labels were revised and sub-categories were created. In this stage, we also used an LLM (ChatGPT, GPT-4o, Aug. \& Sep. 2025 versions) to challenge the categories and labels created (e.g., prompting "How would you label requests that ask for ...?"). This was always only done after creating own meaningful labels or categories based on the data set. At all times, the cognitive and creative lead was held by the human coder. If appropriate, the AI-assisted assessment was adapted to better fit the data and led to further revisions of categories or labels. 

In building and naming intent categories, we followed two principles: Categories should express the task- or action-based function of the expected answer instead of merely being content- or function-based. So, we, for example, named one category "required action(s)" (task-based lens) instead of "dog travel guidelines" (contend-based lens) or "list" (function-based lens). Moreover, to make our taxonomy more applicable to other task contexts, we avoided context-specific, i.e., airport-specific labeling and aimed for more abstract categories. We, for example, labeled a category "scheduled event completion time" instead of "flight arrival time". 
The GT methods of constant comparisons \cite[][p. 187]{charmaz2006constructing} and theoretical comparisons \cite[][pp. 94–95]{corbin2015basics} helped to raise the level of abstraction.

In the final analysis stages, we focused on all mentions that we felt difficult to classify and let these challenge, refine, and detail the taxonomy created so far. 
This has resulted in a \textbf{4-layered taxonomy comprising 20 level-1 question intent categories and 86 sub-categories}.
In Section \ref{sec:findings}, we present this taxonomy in detail.

\subsection{Taxonomy Evaluation}

To verify how well the developed taxonomy categories and their descriptions fit the inquirers' information requests, we gave 147 randomly selected questions (equaling 20 percent of the total number of questions) to an outside assessor not involved in the data collection and analysis of this study, and let the assessor assign the selected questions to the 20 level-1 categories. We made sure that all 20 categories are represented in the sample by randomly selecting questions from each category.

The inter-rater agreement was measured using unweighted Cohen’s $\kappa$ \cite{cohen1960coefficient}. This measure complies with the rating task of assigning each information request to exactly one category \cite[][p. 578]{artstein2008inter}. The agreement between the raters was substantial with a Cohen's Kappa value of $\kappa$ = 0.84, 95\% CI [0.78, 0.91]. This indicates that the developed taxonomy categories and their descriptions are consistent and clear.

It is important to note that measuring inter-rater agreement is not common and mostly inappropriate when using GTM. However, as we aimed at a taxonomy, our results are largely category-based which basically enables us to measure agreement. Only one adjustment was necessary: In our task-based analysis, the inquirers' task context was essential to interpret request intents. So, we made the task context behind each request available to the outside assessor.

\section{Findings: A Taxonomy of Task-Based Information Request Intents}\label{sec:findings}

In the following, we introduce the proposed intent taxonomy and outline its key categories. These key categories represent the most common request intents in our dataset and, at the same time, appear to be the most fruitful ones in terms of applicability to other task contexts (see Section \ref{sec:discussiontransferability} for more details on the transferability of our taxonomy).

\begin{figure}
  \centering
  \includegraphics[width=\linewidth]{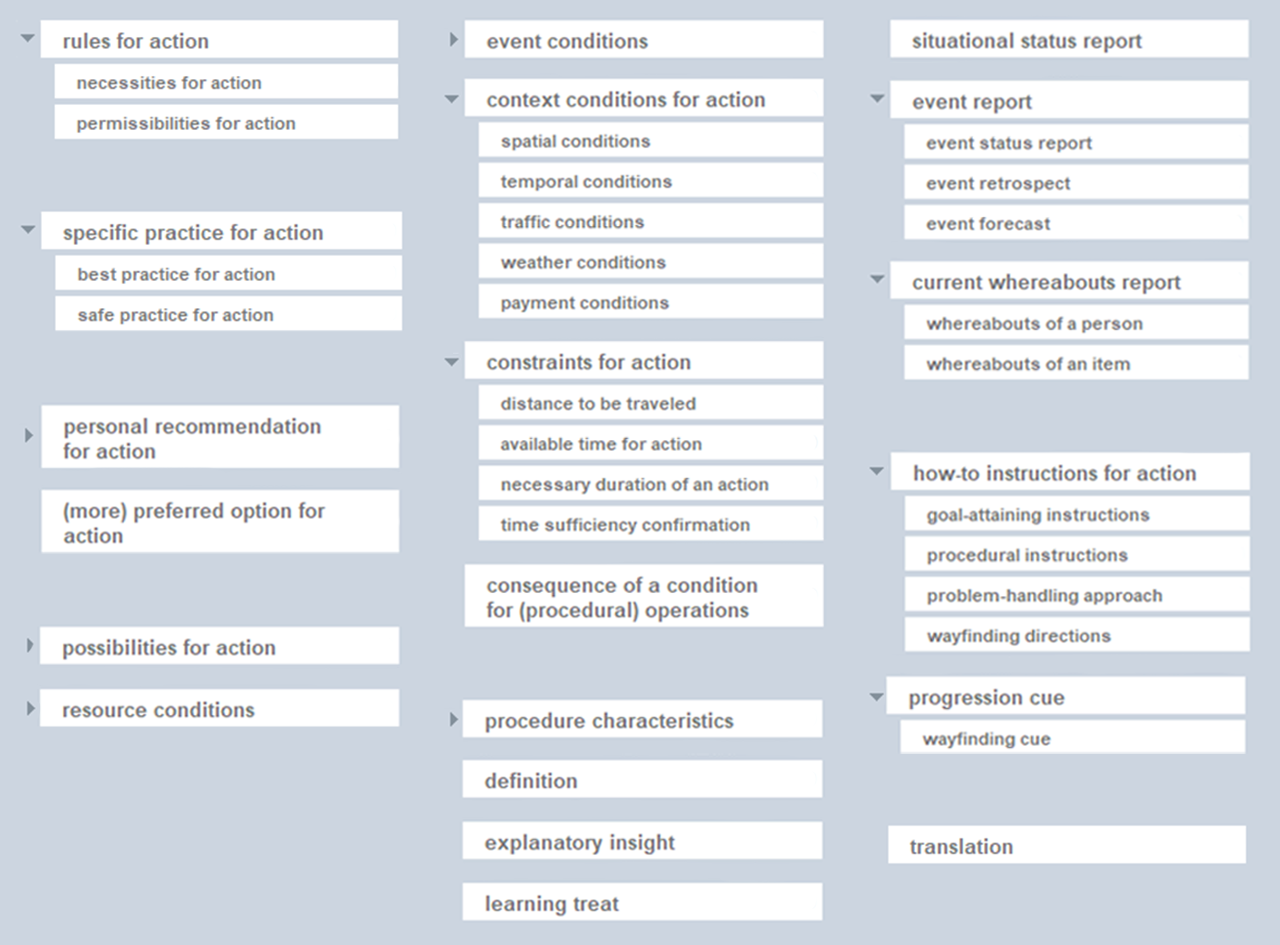}
  \caption{The top levels of the task-based request intent taxonomy}
  \label{fig:taxonomy}
  \vspace{-8mm}
\end{figure}

The top levels of the taxonomy are summarized in Figure \ref{fig:taxonomy}.
An \href{https://osf.io/7crku/overview?view_only=c45d1916f4724522aae967ac896a3e2c}{overview of all intent categories is available on the OSF data repository}. Excerpts from this overview can also be found in the appendix (Tables \ref{tab:actionrules}--\ref{tab:cues}). 
This overview includes descriptions to every category, example requests and associated example tasks, as well as typical patterns of request formulations and pattern variants. With this overview and the following remarks, we aim to provide a comprehensive understanding of our intent classification and support research on (natural-language) information systems for more effective task support.

\subsection{Taxonomy Overview and Distinct Features}

Drawing on 720 questions asked to airport information desk workers, we created a taxonomy with twenty level-1 question intent categories and 86 sub-categories on levels 2--4. 

We found that airport visitors' task-based information seeking is focused on several particular key intents:
For task accomplishment, \textbf{action rules} (e.g., what actions must be performed to attain a particular goal) serve task performers as general guidelines for acting. \textbf{Context conditions} and \textbf{constraints for actions} create a conception of the environment that can help task performers in knowing how to comply with these rules. \textbf{Keeping track of relevant events} is a typical means of inquirers' to be able to adapt their actions to changing conditions.
The surrounding conditions also define the parameters for all other \textbf{action possibilities} and \textbf{specific practices} that inquirers desire to perform. Knowing what possibilities are there helps, for example, in choosing an action. In areas of individual taste, task performers also look for \textbf{personal recommendations on how to act} in order to decide on an action. Here, inquirers with a clearer idea of how events should proceed tend to ask questions about their \textbf{preferred choice for action} instead.
Many questions are also directed toward \textbf{resources}, i.e., useful entities for task performance or achievement, such as services, provisions, and equipment.
When inquirers need help carrying out tasks, they typically look for \textbf{how-to instructions} that provide them with the individual steps necessary to be performed for task completion. People that get stuck 
during task performance seek \textbf{progression cues} helping them to move on. All these key request intents (also: expected answer types) will be explained in more detail below.

\subsection{Rules for Action: Necessities and Permissibilities}

\textbf{Rules for action} concern what must be done (\textbf{necessities}) and what is allowed to be done (\textbf{permissibilities}) in performing a certain task. Inquirers expecting such rules are often seeking this information to plan an action in advance (e.g., "I have asthma. Am I allowed to bring my inhaler on the plane?", but sometimes they also search for such details to immediately act (on it), for example when a traveler at the airport is asking, "I'm flying with [airline]. Where do I need to go?".

Questions about \textbf{necessities} include inquiries about required actions (in the form of \textit{What do I need to do?}), required locations (\textit{Where do I need to do this?}), required means for action (\textit{What means do I need to use here?}), and deadlines for required actions (\textit{By what time do I need to do this?}). Besides this, many inquirers are also interested in whether a certain action needs to be taken (e.g., "Do I have to go through passport control?").

Questions about \textbf{permissibilities} contain comparable inquiries. They concern allowed options (\textit{What am I allowed to do?}), allowed lengths of times (\textit{How long am I allowed to do it?}), and allowed quantities (\textit{How much of this am I allowed?}), as well as different statements of permissibility (\textit{Am I allowed to do this?} or \textit{Am I allowed to do it by this means?}). Table \ref{tab:actionrules} provides concrete example requests to all sub-categories of the \textbf{action rules} category along with typical formulation patterns.

\subsection{Resources and Possibilities for Action}

When asking questions about \textbf{action possibilities}, inquirers explore whether and when certain actions are feasible and which actions are possible. Their aims are diverse: planning a future action, choosing an action out of different options, or carrying out an action immediately. 
When asking about \textbf{possible options for action}, inquirers await a complete list of all options requested. However, interestingly, they typically already limit the requested selection to a specific location, time, or target group in their question. (see Table \ref{tab:possibilities} for more details on action possibilities) 

Requests on \textbf{resource conditions} are closely related to what is expected for action possibilities. Resources, such as services, provisions, and equipment, are useful entities for task performance or achievement. Unsurprisingly, questions regarding the availability of particular resources are, therefore, very diverse (\textit{Is this resource available?}, \textit{Where, when, and how long is it available?},\textit{What resources are available?}) (\href{https://osf.io/7crku/overview?view_only=c45d1916f4724522aae967ac896a3e2c}{find more details on this sub-category on OSF}).

\subsection{Specific Practices for Action}

In the airport context, we found that task performers were interested in different ways of performing actions: Many inquirers asked for \textbf{best practices}, i.e., officially accepted experience reports on what are the most convenient actions, locations, time frames, and means for completing a particular task. 
There are also few requests on \textbf{safe practices}. These concern airline safety and traveling alone. (see Table \ref{tab:specificpractices} for more details on specific practices for action)

\subsection{Recommendations and Preferred Options}

In realms where not prescribed rules and undisputed knowledge, but individual preferences and taste determine goal attainment, task performers expect \textbf{recommendations for actions}. These are personal experience- or opinion-based statements about what actions are good or suitable in a particular situation. We only found requests for recommendations in the domains of consumption, lifestyle, and leisure where inquirers aim for recreation or entertainment (e.g., when they want to go sightseeing or eat out between changing planes). Recommendations always served as a means for choosing an action. However, as opposed to requests asking for possible options for actions, inquirers who wish for recommendations do not pre-select options themselves but expect preselected options. Rotraud (see \href{https://asistdl.onlinelibrary.wiley.com/cms/asset/8a709386-afab-4798-b2c2-e7ffb22a6cd7/asi70010-fig-0001-m.jpg}{Figure 1} in \cite{kilian2025right}) summarizes these expectations as follows: 

 \begin{quote} \small
 "If someone is standing there hungry, and I say, 'Let me show you the menus of all the restaurants around here' […] that’s not what they want." 
 \end{quote}
Nicole understands her role in such situations as "consulting". 
Rotraud further notes that recommendations should best consider inquirers' preferences, for example, recommendations for eating places should only include facilities that reflect the inquirers preferences regarding cuisine and ambiance (see Table \ref{tab:recommendation} for more details on recommendations).

The inquirers' preferences also come into play in situations where they ask for \textbf{a (more) preferred option for action} after an information clerk has suggested a way of performing an action they disfavor (see Table \ref{tab:morepreferredoption} for more details).

\subsection{Conditions and Constraints for Action}

\href{https://osf.io/7crku/overview?view_only=c45d1916f4724522aae967ac896a3e2c}{Tables 7--10 on the OSF data repository} provide an overview of the conditions and constraints that occupy inquirers in their task performances. In our taxonomy, we distinguish between \textbf{context conditions} or \textbf{event conditions} that generally define the scope of inquirers' actions (e.g., weather conditions, traffic conditions) and \textbf{constraints for action} that explicitly restrict inquirers actions (e.g., available time, distance to be traveled). 

Furthermore, airport visitors are interested in \textbf{consequences of conditions for (procedural) operations}, such as effects of an strike on the flight schedule.
\vspace{-1mm}

\subsection{Event Status Reports}

To know how, when, and where to act, airport visitors often need to keep track of events (e.g., checking on one’s relatives' flight arrival time to define one's trip to the airport). \textbf{Event status reports} mostly present updates requested by airport visitors to plan future or upcoming events (\href{https://osf.io/7crku/overview?view_only=c45d1916f4724522aae967ac896a3e2c}{see Table 12 on OSF for more details}). 
\vspace{-1mm}

\subsection{How-to Instructions for Action}\label{subsec:instructions}

Drawing on our data, we can distinguish between four types of how-to instructions:
\textbf{goal-attaining instructions}, \textbf{procedural instructions}, \textbf{problem-handling approaches}, and \textbf{wayfinding directions}. Table \ref{tab:instructions} gives an overview of these sub-categories. We would like to add two notable aspects here:

First, requests for \textbf{goal-attaining instructions}, \textbf{problem-han-dling approaches}, and \textbf{wayfinding directions} often take the form of declarative sentences missing any question (word) (e.g., "I want to take my dog with me to Mallorca", "I missed my flight", and "I need to get to this hotel"). This could be due to the experts' reports being inaccurate. However, it is more likely that these are examples of "'ill-formed' quer[ies]" \cite[][]{ross2019conducting}, a phenomenon that has been referred to frequently in library literature. It describes how library patrons initially do not specify what they want in their queries, or only do so incompletely. In our case, it is incomplete queries. Interestingly, this form recurs repeatedly in our data and seems to be a typical formulation pattern in these sub-categories. 

Second, when we look at typical formulations of wayfinding needs we see that a lot of request formulations follow the pattern \textit{"Where is ...?"}. However, depending on the specific task context, not every \textit{"Where is ...?"} question requires detailed way-finding directions; some simply call for a statement of a location, for example if the inquirer is a more experienced traveler and familiar with the general airport layout. Thus, request intents or expected answer types cannot be derived solely from request formulations.
The experts' experience, however, reveals that task context can help assessing peoples' needs correctly. Nicole, for example, states,

 \begin{quote} \small
"On the phone, you might say, '[...] [The post office] is in Terminal 2, in the arrivals hall.' And that’s fine. They [the inquirers] will think, ‘Alright, it’s in arrivals.’ But once they're actually at the counter, they’ll want to know [...] where exactly [it is.] [...] [T]he directions need to be a bit more specific."
\end{quote}

Thus, during planning tasks, inquirers are typically satisfied with less detailed information on locations. However, when the inquirers are on the verge of performing an action, they need precise instructions if these are missing.
The experts recognize planning tasks based on the inquirers' location (at home vs. on site). The communication channel used (telephone vs. counter) can also serve experts as a proxy for the inquirers' location.
\vspace{-1mm}

\subsection{Progression Cues}

In way-finding contexts, airport visitors sometimes wish for a cue that helps them to move on (e.g., a \textbf{turn instruction} or \textbf{pointer in the right direction}). Such cues are requested in the middle of a task performance. Using GTM comparison methods, we abstracted from these \textbf{wayfinding cues} and grouped them under the more general category of progression cues (see Table \ref{tab:cues} for more details).

\section{Discussion}

In this section, we summarize the main results of our study, followed by a discussion of our task-based intent taxonomy regarding 1) existing classification systems, 2) its applicability to other task contexts, 3)~practical implications, and 4) limitations.
\vspace{-1mm}

\subsection{Main Results}

We contribute a taxonomy of task-based information request intents that bridges the gap between traditional query-focused classifications and the emerging demands of (AI-driven)
task-oriented search. Our taxonomy:

\begin{itemize}
    \item shows \textbf{diverse request intents} that people have while performing tasks in a specific heterogeneous context,
    \item identifies the \textbf{key intents} in this context, and
    \item presents request intents from a \textbf{high-level task perspective} to support generalizability (e.g., "rules for action" instead of  "air travel regulations").
\end{itemize}

Additionally, we found that the level of detail expected for answers is typically lower for planning tasks compared to tasks involving the execution of an action. However, this is not always evident from request formulations (see Section \ref{subsec:instructions} on \textit{"Where is ...?"} questions).
\vspace{-1mm}

\subsection{Comparison with Existing Intent Taxonomies}\label{subsec:compareexisting}

In the following, we compare our task-based intent taxonomy with existing frameworks in QIU\&C and task-based IR. 
This aims to show that our taxonomy is a first step in integrating the two perspectives.

Bu and colleagues' (2010) intent classification for general question answering \cite{bu2010function} adopts a function-oriented viewpoint. It contains six categories: \textit{fact}, \textit{list}, \textit{reason}, \textit{solution}, \textit{definition}, and \textit{navigation}. All of these are reflected in our task-based taxonomy -- four of them one to one (viz., \textit{explanatory insight}, \textit{problem-handling approach}, \textit{definition}, and \textit{wayfinding directions}) and two of them (\textit{fact}, \textit{list}) as implicit intent features (e.g., airport visitors expect facts when they ask for \textit{rules for action} or \textit{resource availabilities} and look for lists when it comes to \textit{possible options for action}.). Cambazoglu and colleagues' (2021)  multi-faceted intent taxonomy for web search \cite{cambazoglu2021intent} contains 16 categories which can be similarly mapped onto our taxonomy. This exemplifies that our taxonomy covers many categories in existing intent classifications (probably due to the diversity of tasks in the airport context). The difference between our taxonomy and existing ones seems not to be a question of scope but one of perspective, viz. task-based versus query-based. This becomes evident in the distinct interpretations of query intents. While \textit{“Where is ...?”} questions, according to \cite{cambazoglu2021intent}, expect a position or location of an entity, we found that, depending on the inquirers' specific task, inquirers also use this question when they expect how-to instructions (viz, wayfinding directions). So, the contextualized perspective of our intent taxonomy may contribute to better understanding user intents in real-world task performances. Eventually, this could add to more effective support in task completion. 

Task frameworks in tasked-based IR describe hierarchies from high-level activities to (work) tasks to search tasks. This way, they advance the understanding of how tasks relate to information seeking. However, these frameworks miss the user lens in terms of search intents. As our taxonomy takes both the user and the task lens, it can function as a first means to integrate task frameworks with request intent taxonomies. This could enhance our understanding of how tasks and information seeking are intertwined. Our taxonomy may, for example, benefit from the (work) role dimension in Soufan and colleagues' (2021) integrated task taxonomy \cite{soufan2021untangling}. Thus far, our taxonomy contains tasks that are motivated by a wide variety of user roles (e.g., inquirers as travelers, as greeters, as parents, as assistants), but there is no dimension that differentiates task performances by user roles. 
\vspace{-2mm}

\subsection{Applicability of the Taxonomy to Other Task Contexts: A Thought Experiment}\label{sec:discussiontransferability}

To develop a realistic taxonomy of tasked-based request intents, it is necessary to start from a specific task context. We have chosen the airport context due to the diverse activities and context conditions that shape airport visitors' information seeking. Because of this diversity, our taxonomy is likely (partly) applicable to other task contexts. In the following, we use a thought experiment combined with existing data from other contexts to explore the scope of the proposed intent taxonomy. Precisely, we examine how well our taxonomy can be applied to the tasks of \textit{cooking} and \textit{accessing public welfare services}.

The context of cooking differs from the airport context in many ways (domestic vs. public, stationary vs. mobile, everyday-life vs. more extraordinary). Nevertheless, many of our intent categories seem to be represented in this setting. This can be seen when mapping cooking-related information needs found in a naturalistic study by \cite{frummet2022what} to our taxonomy: 
Rules for action correspond to recipe instructions and physico-chemical principles in cooking. These may prompt naturalistic requests such as “Do I need to preheat the oven?” and “[H]ow long do I need to bubble [...] the lentils?”.
Resources matter in the cooking context as well (e.g., “Teaspoons are the small ones, right?”). Problem-handling approaches were also found by \cite{frummet2022what} (e.g., “But it has become a bit too watery [...] How can I reduce the liquid?”).
This mapping shows that several of our categories correspond to the cooking context. In some cases (e.g., “[H]ow long do I need to bubble [...] the lentils?”), additional level-3 categories would be necessary (viz., \textit{required} length of time for action) but these would reflect other sub-categories we have built (viz., \textit{allowed} length of time for action).

Information seeking in the public welfare context also has several parallels. \cite{simonsen2020disabled} and \cite{verne2022howcan} examined citizen–chatbot interactions related to public welfare services and report information requests that can be mapped to our taxonomy: 
Rules for action are mirrored in administrative procedures and requirements for completing forms, leading to requests such as “What am I supposed to write in the employment status form when I have paternity leave?”
Resources are also central in this context (e.g., “What is the child benefit this year?”).
Requests for how-to instructions appear as well (e.g., “How do I receive money when I am expecting a child?”).
In some cases (e.g., “What is the child benefit this year?”), sub-categories would need to be expanded (e.g., broadening “resource pricing conditions” to “resource financial conditions” to capture both service costs and monetary benefits).

Taken together, these thought experiments suggest that the proposed intent taxonomy can be basically applied across distinct task contexts. Future work should therefore seek to systematically extend its scope to additional domains by categorizing and integrating request intents present in other settings.
\vspace{-1mm}

\subsection{Practical implications}

By combining user- and task-centered perspectives on human information seeking, the proposed intent taxonomy enables a contextualized, task-based understanding of user intents.
This understanding may, in turn, inform the design of AI-driven conversational search systems (e.g., virtual agents or chatbots) and enhance their task-supportive capabilities. 
To this end, our taxonomy provides developers with a set of tools, including diverse categories of task-based search intents, corresponding request formulation patterns and variants, and relevant contextual conditions (task type). 

Our findings may also be fruitful for evaluating or understanding human-AI interactions from a task perspective. 
\vspace{-1mm}

\subsection{Limitations}
Drawing on expert interview data, our findings have specific empirical limitations. First, our taxonomy only contains request intents that were expressed in interactions with information desk workers. This excludes independent online searches. Nevertheless, our taxonomy covers many categories in existing query-based intent classifications (see Section \ref{subsec:compareexisting}). This may be due to the numerous tasks, fixed workflows, and tight schedules that await airport visitors and prompt them to turn to information desks for help.
Second, the retrospective reports of the experts may show memory errors and biases. We tried to compensate for this limitation by also conducting interviews where the interviewee simultaneously answered incoming information requests.

By building a taxonomy of task-based request intents, we, moreover, focus here on the user perspective in task performance. However, people do not always correctly understand tasks and this can lead to task failure \cite{kilian2025right}. As a remedy, the gathered expert interview data could enrich the proposed taxonomy with additional insights on objective (information) requirements for task completion.

\section{Conclusion}

The proposed taxonomy of task-based request intents bridges the gap between traditional query-focused intent classifications and the emerging demands of AI-driven search.
It takes a task-focused perspective on diverse intents behind users' information requests and links task goals, request formulations, and expected answer types. This can inform the design of search systems toward more effective and user-friendly task support.

\begin{acks}
The authors are very thankful to the interview participants for their time and insights and to the InfoGate Information Systems GmbH for its support in carrying out this research. Special thanks go to Markus Kattenbeck and the student assistants for their help in conducting and transcribing the interviews, respectively.
The authors also thank the anonymous reviewers for their helpful comments and suggestions.

ChatGPT was utilized to revise the language and wording of text passages in this work.
\end{acks}

\bibliographystyle{ACM-Reference-Format}
\bibliography{bibliography}

\appendix


\begin{table*}
\centering
\tiny

\raggedright
\textbf{ \LARGE A\quad APPENDIX \label{sec:appendix} \vspace{+3mm}}

\centering
\caption{"Rules for action" sub-categories}
\label{tab:actionrules}
\vspace{-2mm}

\begin{tblr}{
  cell{2}{1} = {r=6}{}, 
  cell{8}{1} = {r=5}{}, 
  hline{1-2,8,13} = {-}{},
}

\textbf{request intent}                          & \textbf{request sub-intent}                    & {\textbf{intent description}}                & \textbf{associated example task}               & \textbf{example request}                                           & {\textbf{formulation patterns \& variants}}                                                                                                                                                                                                                                                                                                              \\
{\textbf{necessities}\\\textbf{for action}}      & necessity statement                            & {indicates if sth.\\is required}                       & {changing\\planes}                       & {"Do I have to go through\\passport control?"}                     & {\labelitemi\hspace{\dimexpr\labelsep+0.5\tabcolsep}Do I need to do [action]?\\\labelitemi\hspace{\dimexpr\labelsep+0.5\tabcolsep}Must I do~[action]?\\\labelitemi\hspace{\dimexpr\labelsep+0.5\tabcolsep}Do I need~[entity]?}                                                                                                                                     \\
                                                 & required action(s)                             & {indicates\\what
actions\\are required 
}           & {planning an\\air travel\\with a baby}    & {"I'm flying with my baby.\\What do I need to take\\care of?"}     & {\labelitemi\hspace{\dimexpr\labelsep+0.5\tabcolsep}What do I need to do? 
\\\labelitemi\hspace{\dimexpr\labelsep+0.5\tabcolsep}What
do I need to pay attention to?
\\\labelitemi\hspace{\dimexpr\labelsep+0.5\tabcolsep}Which [option] do I need to do?}                                                                                                        \\
                                                 & {required location\\for an action
  }          & {indicates\\where
an action\\must be done}             & {departing\\an airport}                  & {"What’s the gate\\for my flight?"}                                & {\labelitemi\hspace{\dimexpr\labelsep+0.5\tabcolsep}Where
do I need to go?~\\\labelitemi\hspace{\dimexpr\labelsep+0.5\tabcolsep}Where do I need to do [action]?~\\\labelitemi\hspace{\dimexpr\labelsep+0.5\tabcolsep}{[}task description] Is there a [designated location]?\\\labelitemi\hspace{\dimexpr\labelsep+0.5\tabcolsep}What is the~[designated location]?} \\
                                                 & {required location\\confirmation
  }           & {confirms that known\\location is still right}        & {checking\\one's gate}                   & {"My ticket says [gate no.].\\Is that still correct?"}             & {\labelitemi\hspace{\dimexpr\labelsep+0.5\tabcolsep}Is~[designated location]~still correct?\\\labelitemi\hspace{\dimexpr\labelsep+0.5\tabcolsep}It still says [designated location]. Is that still correct?}                                                                                                                                                       \\
                                                 & {deadline\\for a required action
  }           & {latest time by which an\\action must be completed.}  & {planning one's journey\\to the airport} & {"[By w]hat time\\do I need to arrive?"}                           & {\labelitemi\hspace{\dimexpr\labelsep+0.5\tabcolsep}How late can I still do [action]?\\\labelitemi\hspace{\dimexpr\labelsep+0.5\tabcolsep}(By) when do I need to do [action]?}                                                                                                                                                                                      \\
                                                 & {required means\\for
  action
  }              & {indicates what means must\\be used for an action}    & {planning one's\\airport visit}           & {"What [travel] documents\\do I need?"}                            & {\labelitemi\hspace{\dimexpr\labelsep+0.5\tabcolsep}What kinds of [means] do I need?\\\labelitemi\hspace{\dimexpr\labelsep+0.5\tabcolsep}Which[means] do I need to [use] here?}                                                                                                                                                                                     \\
{\textbf{permissibilities}\\\textbf{for action}} & permissibility
  statement                     & {indicates if sth.~\\is allowed}                       & {packing one's\\suitcases}               & {"Am I allowed [...] to bring\\my child’s food~[in my carry-on]?"} & {\labelitemi\hspace{\dimexpr\labelsep+0.5\tabcolsep}Can I/[person] do [action]?\\\labelitemi\hspace{\dimexpr\labelsep+0.5\tabcolsep}Is there any way I can do [action]?\\\labelitemi\hspace{\dimexpr\labelsep+0.5\tabcolsep}May I do [action]?}                                                 \\
                                                 & {means-dependent\\permissibility statement
  } & {indicates if a means\\is allowed}                     & {planning one's\\airport visit}           & {"When I fly to Africa, can\\I travel with just my ID card?"}      & \labelitemi\hspace{\dimexpr\labelsep+0.5\tabcolsep}Can I do [action] by [means]?                                                                                                                                                                                                                                                                                    \\
                                                 & {allowed options\\for
  action
  }             & {indicates~what
actions\\are allowed}                  & {packing one's\\suitcases}               & {"What am I allowed to\\pack in my carry-on bag?"}                 & {\labelitemi\hspace{\dimexpr\labelsep+0.5\tabcolsep}What
can we do?
\\\labelitemi\hspace{\dimexpr\labelsep+0.5\tabcolsep}What
am I allowed to do?~}                                                                                                                                                                                                               \\
                                                 & {allowed length of
  time\\for action
  }      & {indicates how long an action\\is allowed to be done} & {picking so. up\\at the airport}         & {"How long am I allowed\\to park out there?"}                      & {\labelitemi\hspace{\dimexpr\labelsep+0.5\tabcolsep}How long am I allowed to do [action]?\\\labelitemi\hspace{\dimexpr\labelsep+0.5\tabcolsep}How long can I do [action]?}                                                                                                                                                                                         \\
                                                 & {allowed quantity\\for action
  }              & {indicates how much\\of sth. is allowed\\~}            & {packing one's\\suitcases}               & {"How much free luggage\\am I allowed?"}                           & \labelitemi\hspace{\dimexpr\labelsep+0.5\tabcolsep}How much of [entity] am I allowed?                                                                                                                                                                                                                                                                               
\end{tblr}
\end{table*}

\begin{table*}[hb]
\centering
\tiny

\caption{"Specific practices for action" sub-categories}
\label{tab:specificpractices}
\vspace{-2mm}

\begin{tblr}{
  cell{2}{1} = {r=5}{},
  hline{1-2,7-8} = {-}{},
}

\textbf{request intent}                      & \textbf{request sub-intent}             & {\textbf{intent description}}                          & \textbf{associated example task}                 & \textbf{example request}                                                 & {\textbf{formulation patterns \& variants}}                                                                                                                                                                                          \\
{\textbf{best practice}\\\textbf{for action}} & {best
practicability\\statement}           & {indicates~if sth. is\\the best practice 
}                   & {killing time during\\connecting flights} & {"Is it worth going to Munich at~\\all, or should we go to Freising?"}   & {Is it even worth doing [action],\\or should we do [other action]?}                                                                                                                             \\
                                             & {best
option\\for action}               & {indicates what action\\is the most convenient}                  & {choosing a\\travel route}                 & "How can you fly to Mombasa?"                                            & {\labelitemi\hspace{\dimexpr\labelsep+0.5\tabcolsep}How can/do I do [action]?\\\labelitemi\hspace{\dimexpr\labelsep+0.5\tabcolsep}What is the best way to do [action]?} \\
                                             & {best
location\\for an action\textbf{}} & {indicates where it is most\\convenient to perform an action}   & {planning one’s journey\\to the airport}   & "where should I get off?"                                                & {\labelitemi\hspace{\dimexpr\labelsep+0.5\tabcolsep}Should we do [action] here or there?\\\labelitemi\hspace{\dimexpr\labelsep+0.5\tabcolsep}Where should I do [action]?}                                                                        \\
                                             & {best
time (frame)\\for an action}      & {indicates when it is the most~\\convenient time for an action}  & {planning one’s journey\\to the airport}   & {"When is the best time~\\for me to come?"}                              & {\labelitemi\hspace{\dimexpr\labelsep+0.5\tabcolsep}When should I do [action]?\\\labelitemi\hspace{\dimexpr\labelsep+0.5\tabcolsep}When is it best to do [action]?}                                                                              \\
                                             & {best
means\\for an action}             & {indicates what are the most convenient\\means for an action} & {finding one's way\\to the hotel}          & {"I want to go to [hotel]. What is\\the best means of transportation?"} & {[desired action]. What is the best [means\\to attain desire]?}                                                                                                                             \\
{\textbf{safe practice}\\\textbf{for action}} & {(action/resource)\\safety statement}   & {indicates~if sth. is safe}                                 & {finding one's way\\to the hotel}          & "Can
I go there alone?"                                                  & {\labelitemi\hspace{\dimexpr\labelsep+0.5\tabcolsep}Can I do [action] alone?\\\labelitemi\hspace{\dimexpr\labelsep+0.5\tabcolsep}Is [entity] safe?\\} 

\end{tblr}
\end{table*}

\begin{table*}[hb]
\centering
\tiny

\caption{"Personal recommendation for action" sub-categories}
\label{tab:recommendation}
\vspace{-2mm}

\begin{tblr}{
  hlines,
}
\textbf{request intent}                                            & \textbf{intent~description}                                                & \textbf{associated example task}          & \textbf{example request}                                                        & \textbf{formulation patterns~\& variants}                                                                                                                                                                         \\
{\textbf{recommendability}\\\textbf{statement}~}                   & {indicates if an action is\\good/suitable\\(personal
opinion/experience)}  & {killing time during\\connecting flights} & {"Is it worth going there [tourist\\destination]? What would you\\recommend?"~} & {\labelitemi\hspace{\dimexpr\labelsep+0.5\tabcolsep}Is it worth doing [action]? What would you recommend?~\\\labelitemi\hspace{\dimexpr\labelsep+0.5\tabcolsep}Does it pay off to do [action]?~\\\labelitemi\hspace{\dimexpr\labelsep+0.5\tabcolsep}{[}service offer]. Are you familiar with this?}                                                                                                                                                                                                               \\
{\textbf{recommended options}\\\textbf{for action}}                 & {indicates what action(s) are\\good/suitable (opinion)}                    & {satisfying one's\\hunger}                & "What's good here at the airport?"                                              & {\labelitemi\hspace{\dimexpr\labelsep+0.5\tabcolsep}What
should I do? What would you recommend?~\\\labelitemi\hspace{\dimexpr\labelsep+0.5\tabcolsep}We'd like to do [action]. What can you recommend?~\\\labelitemi\hspace{\dimexpr\labelsep+0.5\tabcolsep}{[}service offer] What would you recommend?/What is good?\\\labelitemi\hspace{\dimexpr\labelsep+0.5\tabcolsep}What can you recommend in [location]?~\\\labelitemi\hspace{\dimexpr\labelsep+0.5\tabcolsep}Do you know [entity]? Do you have any tips?} \\
{\textbf{preference-dependent}\\\textbf{recommended location(s) }} & {indicates where which location is\\good/suitable for an action (opinion)} & {killing time during\\connecting flights} & "Where
is it worth going?"                                                      & \labelitemi\hspace{\dimexpr\labelsep+0.5\tabcolsep}Where
is it worth going?                                                               
\end{tblr}

\end{table*}

\begin{table*}[hb]
\centering
\tiny

\caption{"(More) preferred option for action" category}
\label{tab:morepreferredoption}
\vspace{-2mm}

\begin{tblr}{
  hlines,
}
\textbf{intent~description}                                                                                               & \textbf{associated example task} & \textbf{example request}                                                             & \textbf{formulation patterns~\& variants}                                                                                                                                                                                                                            \\
{indicates a way of performing an action that is
  (more)\\preferred by the inquirer than an action previously suggested} & planning one’s airport visit     & {"Can’t I just start a sightseeing tour\\{[}...] directly from here~[the airport]?"} & {\labelitemi\hspace{\dimexpr\labelsep+0.5\tabcolsep}Can’t I just do [action]?\\\labelitemi\hspace{\dimexpr\labelsep+0.5\tabcolsep}Isn’t there a [resource] that provides [service]?\\\labelitemi\hspace{\dimexpr\labelsep+0.5\tabcolsep}Is there another~option?“~} 
\end{tblr}

\end{table*}

\begin{table*}[hb]
\centering
\tiny

\caption{"Possibilities for action" sub-categories}
\label{tab:possibilities}
\vspace{-2mm}

\begin{tblr}{
  cell{4}{1} = {r=3}{},
  hline{1-4,7-8} = {-}{},
}
\textbf{request intent}                                    & \textbf{request sub-intent}                          & \textbf{intent~description}                                          & \textbf{associated example task}          & \textbf{example request}                                                             & \textbf{formulation patterns~\& variants}                                                                                                                                                                                                                                                                                                                                                                                                            \\
\textbf{feasibility statement}                             & -                                                    & {indicates~if an action\\is feasible}                                & {reacting to\\adverse conditions}         & {"Can we rebook\\{[}our flight]?"~}                                                  & {\labelitemi\hspace{\dimexpr\labelsep+0.5\tabcolsep}Can I do [action] (here)?\\\labelitemi\hspace{\dimexpr\labelsep+0.5\tabcolsep}Is it possible to do [action] (here)?\\\labelitemi\hspace{\dimexpr\labelsep+0.5\tabcolsep}Do I have the option of doing [action] (here)? ~\\\labelitemi\hspace{\dimexpr\labelsep+0.5\tabcolsep}{[}Condition].Can I do [action]?\\\labelitemi\hspace{\dimexpr\labelsep+0.5\tabcolsep}Can I already/still do [action]?} \\
{\textbf{means-dependent}\\\textbf{feasibility statement}} & -                                                    & {indicates if an action is feasible\\with a particular means}        & {finding one’s way\\to the hotel}         & {"Can I get there by\\public transport?"}                                            & {\labelitemi\hspace{\dimexpr\labelsep+0.5\tabcolsep}Can I achieve [goal] by [means]?\\\labelitemi\hspace{\dimexpr\labelsep+0.5\tabcolsep}Can I do [action] by [means]?~ ~~}                                                                                                                                                                                                                                                                           \\
{\textbf{(possible) options}\\\textbf{for action}}          & {(possible)\\actor-dependent options\\for action}    & {indicates~what actions are feasible 
\\by/with a particular actor} & {spending one's day\\at the airport}      & {"What can we do with\\children?"}                                                   & {\labelitemi\hspace{\dimexpr\labelsep+0.5\tabcolsep}We have [actor] with us. What can we do?\\\labelitemi\hspace{\dimexpr\labelsep+0.5\tabcolsep}What can we do with [actor]?}                                                                                                                                                                                                                                                                       \\
                                                           & {(possible)\\location-dependent options\\for action} & {indicates~what actions are feasible 
\\in a particular place}      & {killing time during\\connecting flights} & "What can we do around here?"                                                        & {\labelitemi\hspace{\dimexpr\labelsep+0.5\tabcolsep}What can I do now? I’ve got [amount of time].\\\labelitemi\hspace{\dimexpr\labelsep+0.5\tabcolsep}What can I do? I’ve got [amount of time] right now.\\\labelitemi\hspace{\dimexpr\labelsep+0.5\tabcolsep}I've got [amount of time]. What can I do during that time?\\\labelitemi\hspace{\dimexpr\labelsep+0.5\tabcolsep}I have some time now. What can I do in that case?}                      \\
                                                           & {(possible)\\time-dependent options\\for action}     & {indicates~what actions are feasible\\at a particular
time}          & {killing time during\\connecting flights} & {"I have, uh, a ten-hour layover\\at the airport. What can I do\\during that time?"} & {\labelitemi\hspace{\dimexpr\labelsep+0.5\tabcolsep}What can I do in [location]?\\\labelitemi\hspace{\dimexpr\labelsep+0.5\tabcolsep}What is there to do in [location]?\\\labelitemi\hspace{\dimexpr\labelsep+0.5\tabcolsep}I'd like to go to [location]. What can I do there?}                                                                                                                                                                      \\
{\textbf{possible times}\\\textbf{for action}}              & -                                                    & {indicates when an action\\is feasible~}                             & planning a trip to a city                 & {"[I’d] like to fly to Paris and\\get the [flight] times."}                          & \labelitemi\hspace{\dimexpr\labelsep+0.5\tabcolsep}I'd like to [action] and get the times~[for action].                                                                                                                                                     
\end{tblr}

\end{table*}

\begin{table*}[hb]
\centering
\tiny

\caption{"How-to instructions for action" sub-categories}
\label{tab:instructions}
\vspace{-2mm}

\begin{tblr}{
  hlines,
}
\textbf{request intent}              & \textbf{intent~description}                                  & \textbf{associated example task} & \textbf{example request}                          & \textbf{formulation patterns~\& variants}                                                                                                                                                                                                                                                                                                                                                                                         \\
\textbf{goal-attaining instructions} & {(step-by-step) instructions on how\\to attain a goal set}      &                                  & "[I] want to eat Leberkäse."                      & \labelitemi\hspace{\dimexpr\labelsep+0.5\tabcolsep}I want/would like to do [action].                                                                                                                                                                                                                                                                                                                                              \\
\textbf{procedural instructions}     & {(step-by-step) instructions on how to\\carry out a procedure}  &                                  & "I want to travel with my dog. How do I do that?" & \labelitemi\hspace{\dimexpr\labelsep+0.5\tabcolsep}I want to do [action].~(How do I do that?)                                                                                                                                                                                                                                                                                                                                      \\
\textbf{problem-handling approach}    & {(step-by-step) instructions on how\\to handle/solve a problem} & departing an airport             & "I missed my flight."                             & {\labelitemi\hspace{\dimexpr\labelsep+0.5\tabcolsep}{[}problem] (What do/should I do now/next? / What now?)\\\labelitemi\hspace{\dimexpr\labelsep+0.5\tabcolsep}What should I do when [problem]?\\\labelitemi\hspace{\dimexpr\labelsep+0.5\tabcolsep}How can I achieve [goal] (now) (when [problem])?}                                                                                                                            \\
\textbf{wayfinding directions }       & {(step-by-step) instructions on how to\\find one's way}         & leaving an airport               & "How do I get to [specific hotel]?"               & {\labelitemi\hspace{\dimexpr\labelsep+0.5\tabcolsep}How do I get to [location]?\\\labelitemi\hspace{\dimexpr\labelsep+0.5\tabcolsep}Where is [location]?\\\labelitemi\hspace{\dimexpr\labelsep+0.5\tabcolsep}How/Where do I find [location]?\\\labelitemi\hspace{\dimexpr\labelsep+0.5\tabcolsep}I'd like to find/I'm looking for [location].\\\labelitemi\hspace{\dimexpr\labelsep+0.5\tabcolsep}I need (to get to) [location].} 
\end{tblr}

\end{table*}

\begin{table*}[hb]
\centering
\tiny

\caption{"Progression cue" sub-categories}
\label{tab:cues}
\vspace{-2mm}

\begin{tblr}{
  cell{2}{1} = {r=4}{},
  hline{1-2,6} = {-}{},
}
\textbf{request intent}                                                                 & \textbf{request sub-intent} & \textbf{intent~description}                                                      & \textbf{associated example task} & \textbf{example request}              & \textbf{formulation patterns~\& variants}                                                                                                                                                                                               \\
{\textbf{wayfinding cue}\\(a cue that helps a\\person to move on\\with finding their\\way)} & route
confirmation          & confirms that one is still on the right way                                      & finding one’s way                & "Am I still [going the] right~[way]?" & \labelitemi\hspace{\dimexpr\labelsep+0.5\tabcolsep}Am I still right?                                                                                                                                                                    \\
                                                                                        & direction
pointer           & {indicates/displays the direction\\to a location within sight}                   & finding one’s way                & "Where [which direction] now?"        & \labelitemi\hspace{\dimexpr\labelsep+0.5\tabcolsep}Where now?                                                                                                                                                                           \\
                                                                                        & turn
instruction            & indicates where to turn                                                          & finding one’s way                & "Is that
terminal 1 or 2?"    & {\labelitemi\hspace{\dimexpr\labelsep+0.5\tabcolsep}Do I need to go [path] or [other path]?\\\labelitemi\hspace{\dimexpr\labelsep+0.5\tabcolsep}Is that [location on road sign] or [location on road sign]?~}                     \\
                                                                                        & cognitive
mapping cue       & {clarifies received wayfinding directions\\to make them fit one’s cognitive map} & finding one’s way                & "Do you mean my right or your right?" & {\labelitemi\hspace{\dimexpr\labelsep+0.5\tabcolsep}Do you mean~[wayfinding entity] or [wayfinding entity]?\\\labelitemi\hspace{\dimexpr\labelsep+0.5\tabcolsep}Is that [wayfinding entity] or [wayfinding entity]?~\\~} 
\end{tblr}

\end{table*}

\end{document}